\documentclass[aps,reprint,superscriptaddress,amsmath,amssymb,amsfonts,prb]{revtex4-1}
\usepackage{xcolor}
\definecolor{darkgreen}{rgb}{0,0.6,0}
\definecolor{cyan}{rgb}{0,0.7,0.8}
\usepackage[colorlinks=true,citecolor=darkgreen,urlcolor=cyan]{hyperref}

\usepackage{graphicx}
\usepackage[caption=false]{subfig}
\usepackage{braket}
\usepackage{epstopdf} 
\newcommand{\rcite}[1]{Ref.~\onlinecite{#1}}

\begin{document}

%\preprint{APS/123-QED}

% Suggested titles of paper:
%\title{Phase diagram of Interacting, and Braiding Ising Anyons on a two-leg ladder} 
%\title{Phase diagrams of non-Abelian anyonic Hubbard model}
% \title{Braid statistics dependent phase transitions in anyonic systems}
\title{Phase transitions on a ladder of braided non-Abelian anyons}

% Force line breaks with \\

\author{Babatunde M. Ayeni}
\email{babatunde.ayeni@mq.edu.au}
\affiliation{Centre for Engineered Quantum Systems, Department of Physics \& Astronomy, Macquarie University, NSW 2109, Australia}
\author{Robert N. C. Pfeifer}
\affiliation{Department of Physics \& Astronomy, Macquarie University, Sydney, NSW 2109, Australia}
%\author{Nathan McMahon}
%\affiliation{Center for Engineered Quantum Systems, School of Mathematics \& Physics, The University of Queensland, St Lucia, Queensland 4072, Australia}
\author{Gavin K. Brennen}
\affiliation{Centre for Engineered Quantum Systems, Department of Physics \& Astronomy, Macquarie University, NSW 2109, Australia}

%\altaffiliation[Also at ]{Centre for Engineered Quantum Systems, Dept. of Physics \& Astronomy, Macquarie University, NSW 2109, Australia.}
%\affiliation{%

\date{\today}% It is always \today, today,

\pacs{Valid PACS appear here}% PACS, the Physics and Astronomy

\begin{abstract}
Non-Abelian anyons can exist as point-like particles in two-dimensional systems, and have particle exchange statistics which are neither bosonic nor fermionic. Like in spin systems, the role of fusion (Heisenberg-like) interactions between anyons has been well studied. However, unlike our understanding of the role of bosonic and fermionic statistics in the formation of different quantum phases of matter, little is known concerning the effect of non-Abelian anyonic statistics. We explore this physics using an anyonic Hubbard model on a two-legged ladder which includes braiding and nearest neighbour Heisenberg interactions among anyons. We study two of the most prominent non-Abelian anyon models: the Fibonacci and Ising type. We discover rich phase diagrams for both anyon models, and show the different roles of their fusion and braid statistics.
\end{abstract}

\maketitle
%\tableofcontents

\section{Introduction}
In two dimensions pointlike particles named anyons which cannot be classified as either bosons or fermions can exist. Unlike bosons or fermions which acquire a  phase factor of $+1$ or $-1$ under exchange, anyons can acquire any complex number phase factor for particles which are classified as Abelian, or a matrix valued action on a set of degenerate states for those classified as non-Abelian. 

Theoretical proposals for systems which may support anyons as emergent quasiparticles include fractional quantum Hall systems and two-dimensional spin liquids,~\citep{laughlin1983,halperin1984,fradkin1989,read1999,read2000,xia2004,Stern2007,Nayak2008,pan2008,stern2010,sanghun-an2011,clarke2013,mong2014,vaezi2014} one dimensional nanowires,\cite{kitaev2001,stanescu2013,klinovaja2014,Nadj-perge2014} and ultra-cold atoms in optical lattices.\cite{Pachos2012} Experimentally, the recent evidence for Majorana edge modes (i.e. Ising anyons) might be closing the gap between theory and experimental realisation of anyons. Aside from the inherent theoretical interest in their exotic properties, there is an ongoing experimental effort to produce these particles in the laboratory for the purpose of fault tolerant topological quantum computing.~\cite{Nayak2008,kitaev2003,freedman2002}

While a solid body of theoretical work has been developed to explain the emergence of anyons from non-anyonic microscopic physics, \cite{Nayak2008} relatively little has been done on the physics of anyons themselves. An important realization was the discovery that interacting anyons in one dimension map onto well known conformal field theories (CFT).\cite{Feiguin2007} With evidences mounting for the ability to engineer quantum systems with anyonic excitations, it becomes imperative to understand their braid and fusion statistics in higher dimensions.

The formation of different possible phases in a collective system of particles are dictated by various competing terms of the system, including local interaction between particles, hopping of particles between sites, the dimensionality and topology of the system, geometric frustrations, and particle statistics. For a review of these concepts, see these Refs.~\onlinecite{Bloch2008,Cazalilla2011,Giamarchi2004}. 

Several anyon models have been studied by different authors including, one dimensional and quasi-1D chains of static SU(2)$_k$ anyons with local antiferromagnetic Heisenberg-like interaction.\cite{Feiguin2007,Trebst2008, Trebst2008a, Poilblanc2011a,Poilblanc2011} In Refs.~\onlinecite{Poilblanc2011a,Poilblanc2013} the authors mapped out the phase diagram for  chains of itinerant and interacting anyons. However, because of the restricted dimensionality, none of those models allow for braiding of anyons. 

Some recent work have been done on braided anyons. In \rcite{Soni2016}, the authors studied two and three leg ladders of Fibonacci anyons, where the anyons could interact and braid with one another. There, the focus was on the strong rung coupling regime where the rung hopping is greater than the leg hoping and their results showed that those models can be essentially mapped onto one dimensional physics of itinerant hard-core anyons. That investigation used exact diagonalization and the results were limited to ladder sizes of ~20 rungs.

Until recently, one of the obstacles to a more complete study of phases of non-Abelian anyons is the lack of good numerical techniques, particularly since the Hilbert space for fusion outcomes of many anyons grows exponentially with the number of anyons, and also does not decompose into a tensor product structure. To address this gap, in a series of research effort, we developed a suite of numerical algorithms based on tensor networks to simulate the physics of non-Abelian anyons in finite or infinite quasi-1D systems including, anyonic Time Evolving Block Decimation (TEBD), \cite{Singh} anyonic$\times$U(1) Matrix Product States (MPS),\citep{Ayeni2015} and anyonic Density Matrix Renormalization Group (DMRG).\cite{Pfeifer2015} In Ref.~\onlinecite{Ayeni2015}, we studied a specific case of braided Fibonacci anyons on an infinite two-leg ladder and among other results, we numerically showed that particle-hole duality breaks down for non-Abelian anyons as would be expected. In Ref.~\onlinecite{Pfeifer2015a}, one of us studied a ladder of $\mathbb{Z}_{3}$ anyons, and found phases with normal and superfluid behaviours. In both of these works, braid statistics was shown to affect, and perhaps, even induce the observed phases.

 % What we do in this work
In this work, we study the physics of both Fibonacci and Ising anyons on a two-legged ladder system over a large range of tunneling couplings and chemical potentials, and fixed interaction coupling. A two-leg ladder is a minimal geometry that allows for particle exchanges, and may provide intuition for the physics of fully two dimensional systems of anyons. Each site on the ladder can either be vacant or has a single non-trivial anyonic charge. Particles on nearest neighbouring sites are allowed  to interact and braid around each other. In analogy with fermionic and bosonic Hubbard model, we refer to this as an \emph{anyonic Hubbard model} on a ladder. By comparing the phase diagrams of Fibonacci and Ising anyons---two of the most prominent species, we establish that despite the microscropic Hamiltonians for the two anyon models being \emph{similar},\footnote{Similarity as used here means that, except that the particle content on a ladder of Fibonacci and Ising anyons are different, the way the particles are coupled are similar. As an analogy, a system of spinless fermions and hardcore bosons only differ in statistics, otherwise they are ``similar''} their  phase diagrams differ.  We conclude that fusion (and braid) statistics certainly play a role in the formation of phases of anyons.

% Organization of sections of the paper 
The rest of the paper are organised as follows: In Section \ref{Sec:Description-of-model}, we introduce and describe our model. In Section \ref{Sec:Results}, we present the results of our work, including the  phase diagrams of the two anyon models, the plots of the entanglement entropy, and the central charges of the critical phases in our models. In Section \ref{Sec:Discussion}, we discuss the results. We conclude in Sec. \ref{Sec:Conclusion} with a summary and outlook.

%%% Description of the model
\section{Anyonic Hubbard model on a ladder}
\label{Sec:Description-of-model}
\subsection{Model and its Hilbert space}
\begin{figure}
\includegraphics[width=1.0\linewidth]{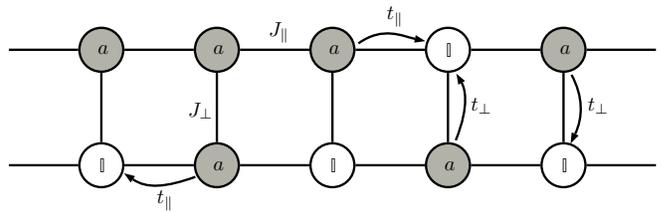}
\caption{A two-leg ladder of interacting and itinerant non-Abelian anyons. An empty site is equivalent to a vacuum charge. When a site is occupied with a charge $a$, the charge can be either a Fibonacci anyon $\tau$ or an Ising anyon $\sigma$ depending on the model studied. (The fermion $\psi$ in the particle spectrum $\mathcal{A}^{\mathrm{Ising}}$ will not be physically supported on sites). 
%Particles on nearest neighbouring sites interact with an Heisenberg-like couplings, with $J_{\perp}$ on the rung, and $J_{\parallel}$ on the legs. When there is a vacant neighbouring site, the anyon with charge $a$ can hop to the empty site, on the legs with an hopping strength $t_{\parallel}$, and on the rungs with $t_{\perp}$. The hopping of anyons on this minimal geometry allows for braiding.}
}
\label{IsingAnyonsLadder}
\end{figure}

% General introduction to the ladder, and its diagrammatic depiction
The system is a two-leg ladder of interacting and braiding non-Abelian anyons. The ladder has a finite vertical width of two and an infinite horizontal dimension (see Fig.~\ref{IsingAnyonsLadder}). There is an imposed ``hardcore'' constraint on each site to penalise the occupation of more than one particle. The specific particles supported are chosen from the Fibonacci and Ising anyons particle spectra: $\mathcal{A}^{\mathrm{Fib}} = \{ \mathbb{I}, \tau \}$, and $\mathcal{A}^{\mathrm{Ising}} = \{ \mathbb{I}, \sigma, \psi \}$, where $\mathbb{I}$ represent the vacuum charge (or empty site), $\tau$ is one Fibonacci anyon, $\sigma$ is one Ising anyon, and $\psi$ is a fermion. All the particles except the fermion $\psi$ are allowed on site, since here it is treated not as a fundamental particle but rather a composition of two $\sigma$ particles. Such a treatment is appropriate for physical systems where direct creation of fermions out of the vacuum is not allowed or is a higher energy process. The fusion algebra and other data of these particles are catalogued in Appendix~\ref{App:FusionData}. We study models with each anyon species independently, and not an eclectic mix of both Ising and Fibonacci anyons.

\begin{figure}
\includegraphics[width=1.0\linewidth]{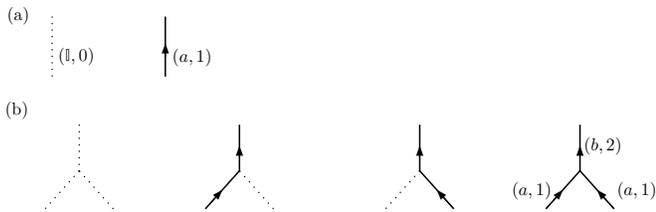}
\caption{Diagrammatic representations of anyonic charges. $(a)$ Two types of lines representing the two types of charges permitted on each site: a vacuum charge $(\mathbb{I},0)$ and a non-trivial charge $(a,1)$, where $a = \tau \text{ or } \sigma$. $(b)$ The charge fusion outcomes on each rung, where the lines carry the same meaning as in $(a)$. The charge label  $b$ in the last vertex diagram can be, $b = \mathbb{I}$ if $a$  is either a $\tau$ or $\sigma$---as both can annihilate to vacuum, or $b=\tau$ if $a=\tau$, or $b=\psi$ if $a=\sigma$---all based on the fusion rules of the individual anyon model. Note that, as Fibonacci and Ising anyons are self dual charges, lines do not need arrowheads, but in the presence of a U(1) number charge---which is not self-dual, the lines need arrowheads.}
\label{Fig:ChargesRepresentation}
\end{figure}

% The anyonic charges ad their diagrammatix representations
The charges in an anyonic system can be labeled in different ways depending on the types of conserved charges in the system. In this work, we find it convenient to either label anyons explicitly by $a$, when the only conserved charges in the Hamiltonian are anyonic, where $a$ is the anyonic charge, or by a composite charge label $(a,n)$, when the Hamiltonian has a U(1) particle number symmetry, where $n$ now represents the number of anyonic charges on a site. The $(a,n)$ composite charge label can also be re-written as $a_n$. For example,  we associate a number $n=0$ to the vacuum charge $\mathbb{I}$, and $n=1$ to either one Fibonacci ($\tau$) or one Ising anyon $(\sigma)$.  Table~\ref{Tb:OnsiteCharges} summarises our charge enumeration conventions.
%\newpage
\begin{table}
\begin{ruledtabular}
\centering
\begin{tabular}{  c    c  } 
Fibonacci anyons   &   Ising anyons \\
\hline \\
$ a = 
\begin{cases} 
\mathbb{I} \equiv (\mathbb{I}, 0) \equiv \mathbb{I}_0   \\  
\tau  \equiv (\tau, 1)  \equiv \tau_1
\end{cases}
$
& 
$ a = 
\begin{cases}
 \mathbb{I}  \equiv (\mathbb{I}, 0) = \mathbb{I}_0  \\ 
 \sigma \equiv (\sigma, 1) \equiv \sigma_1
 \end{cases} 
 $
\end{tabular}
\end{ruledtabular}
\caption{Table shows onsite charges and our different enumeration conventions. The vacuum charge is written as $ \mathbb{I} \equiv (\mathbb{I},0) \equiv \mathbb{I}_0$. One Fibonacci and Ising anyon is written as $\tau \equiv (\tau,1) \equiv \tau_1$  and $\sigma \equiv (\sigma,1) \equiv \sigma_1$. }
\label{Tb:OnsiteCharges}
\end{table} 
As the ladder has a finite vertical width, it sometimes proves convenient to view it as a one dimensional system by coarse-graining the rungs to a single site, thereby collapsing the ladder into a chain. The charges on each sites of the chain would be the fusion outcomes of the charges on the rungs of the original ladder. Table~\ref{Tb:ChargesOnRung} lists the combined charges on the rungs.
\begin{table}
\begin{ruledtabular}
\centering
\begin{tabular}{  c    c  } 
Fibonacci anyons   &   Ising anyons \\
\hline \\
$ (\mathbb{I}, 0) \times (\mathbb{I}, 0) = (\mathbb{I}, 0) \equiv \mathbb{I}_0$ & $ (\mathbb{I}, 0) \times (\mathbb{I}, 0) = (\mathbb{I}, 0) \equiv \mathbb{I}_0$ \\
$ (\mathbb{I}, 0) \times (\tau, 1) = (\tau, 1) \equiv \tau_1$ & $(\mathbb{I}, 0) \times (\sigma, 1) = (\sigma, 1) \equiv \sigma_1$ \\
$ (\tau, 1) \times (\mathbb{I}, 0)  = (\tau, 1) \equiv \tau_1$  &  $ (\sigma, 1) \times (\mathbb{I}, 0) =  (\sigma, 1) \equiv \sigma_1$ \\
$ (\tau,1) \times (\tau,1) = 
\begin{cases} 
(\mathbb{I},2) \equiv \mathbb{I}_2 \\
(\tau,2) \equiv \tau_2 \\
\end{cases} $  & 
$ (\sigma,1) \times (\sigma,1) = 
\begin{cases} 
(\mathbb{I},2) \equiv \mathbb{I}_2 \\
(\psi,2) \equiv \psi_2 \\
\end{cases}
$
\end{tabular}
\end{ruledtabular}
\caption{Table shows the combined total charges on a single rung, derived from fusing the onsite charges of the two sites on that rung. In the fusion $(a_1,n_1) \times (a_2, n_2)$, the left ordered and right ordered pair represent the charge on the top site and bottom site of the rung, respectively. Fusion of the anyonic charges follow the fusion rules of the  specific anyon model (see Appendix~\ref{App:FusionData}), and the number charges fuse using ordinary addition. As in Table~\ref{Tb:OnsiteCharges}, the subscript $n$ in $a_n$ is the number of non-trivial charges (e.g. $\tau$ or $\sigma$) fusing into the anyon charge $a$.}
\label{Tb:ChargesOnRung}
\end{table}
In addition, we represent the anyonic charges on sites and their fusion outcomes on the rungs using a set of diagrammatic representations shown in Fig.~\ref{Fig:ChargesRepresentation}.

% Hilbert space, symmetries and all that
The Hilbert space of our model admits both anyonic and U(1) symmetry. The U(1) symmetry is defined in terms of conserved global particle density. At any fixed global particle number $N$, particles can have different spatial configurations, which are specified by a set of configuration $\mathcal{I}_N$. As the particles are non-Abelian anyons, they also have a fusion space when $N > 0$. The Hilbert space structure is therefore a direct sum, 
\begin{equation}
\displaystyle{\bigoplus_{i \in \mathcal{I}_N}} \mathbb{V}_i^{(N)},
\end{equation} 
where $i$ is the index of the spatial configurations of the particles and $\mathbb{V}_i^{(N)}$ is the anyonic fusion space of that spatial configuration. For all possible particle numbers $N$ of the system, the total Hilbert space is therefore, 
\begin{equation}
{V} = \displaystyle{\bigoplus_{N = 0}^{\infty}}~ \displaystyle{\bigoplus_{i \in \mathcal{I}_N}} \mathbb{V}_i^{(N)}.
\end{equation}

\subsection{Hamiltonian}
We construct a Hamiltonian $\hat{H}$, similar to the Hubbard model, for the interaction and dynamics  of particles on the ladder (see Fig.~\ref{IsingAnyonsLadder}). This consists of a kinetic and chemical potential term, jointly represented by  $\hat{H}_{\text{t-}\mu}$, and an interaction term $\hat{H}_{\text{J}}$, so that 
\begin{equation}
\hat{H} = \hat{H}_{\text{t-}\mu} + \hat{H}_{\text{J}}. 
\end{equation} 
As with fermions and bosons, the kinetic term accounts for the hopping of particles between sites, but now in addition, because the particles are anyons, they also braid when they hop past one another. The chemical potential term control the filling of the ladder. The interaction term acts locally between nearest-neighbouring anyons. The fusion outcome of two Fibonacci anyons is $\tau \times \tau \rightarrow \mathbb{I} + \tau$, and that of two Ising anyons is $\sigma \times \sigma \rightarrow \mathbb{I} + \psi$. We chose the anti-ferromagnetic Heisenberg interaction which assigns lowest energy to the projection of a pair of non-Abelian anyons to the vacuum charge sector, in analogy to favouring the fusion of two spin-$1/2$ particles into a singlet state.

\begin{figure}
 \includegraphics[width=1.0\linewidth]{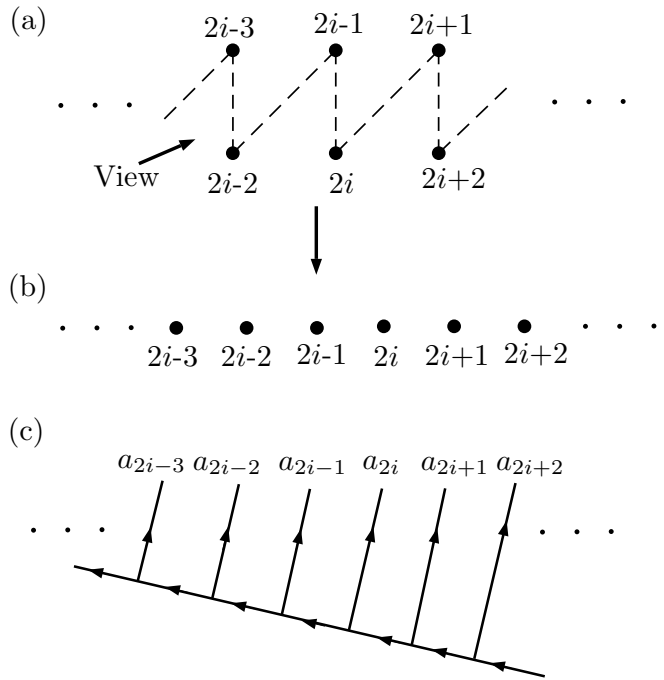}
\caption{(a) Orientation of ladder: The ladder is viewed at an angled position from the bottom to top. Particles on the top leg are regarded to be ``behind'' those on the bottom leg. (b) Linear ordering: The ladder is projected to one dimension; the ladder is ``stretched out" into a chain. (c) Fusion ordering: The particles on the sites of the ladder are fused in the usual way---from left to right. The charge label of site $i$ is denoted as $a_i$ (which can be a  vacuum charge or a non-trivial charge). The intermediate fusion outcomes on the links of the fusion tree have been suppressed for simplicity.}
\label{FusionOrder}
\end{figure}

To capture the physics of braiding, it is necessary to map the sites of the two dimensional ladder to a canonical ordering in one dimension which can then be used to order charges on the leaves of a fusion diagram.  Using Fig.~\ref{FusionOrder} as a reference, the ladder should be viewed from bottom to top, where the particles on the top leg of the ladder are assumed to be ``behind'' those on the bottom leg, and we adopt a zig-zag linear ordering of the sites, which minimizes the interaction length the most. The anyons are fused in the ``standard'' convention: from left to right. 

\subsubsection{Kinetic and chemical potential term}
We derive the contributions to the kinetic energy and chemical potential term using Fig.~\ref{FusionOrder}(i). The kinetic term of the Hamiltonian receives contributions from a possible combination of the following three scenarios: $(1)$ the hopping of a non-trivial charge between sites $2i-1$ and $2i+1$ on the top leg, while passing behind site $2i$ on the bottom leg, and braiding with any non-trivial charge there, otherwise, the passage does not introduce any braid factor into the state of the system. $(2)$ Similarly, the hopping of a non-trivial charge between sites $2i$ and $2i+2$ on the bottom leg, while passing in front of site $2i+1$ on the top leg, and also braiding with any non-trivial charge there, otherwise that passage also does not introduce any braid factor into the state of the system. $(3)$ And lastly, the hopping of a charge between sites $2i-1$ and $2i$ on the rung, which does not introduce any braid factor---as they are nearest neighbouring sites based on our choice of linear ordering. Note that, because of the hardcore constraint, there needs to be a vacant site for hopping to take place. 
The chemical potential is the energy needed to add or remove particles from the sites of the ladder. 

In the convenient diagrammatic formalism similar to those used in Refs.~\onlinecite{bonderson2008, 
Bonderson2007}, the kinetic and chemical potential term, $\hat{H}_{\mathrm{t}\text{-}\mathrm{\mu}}$, is, 
\begin{equation}
 \includegraphics[width=\linewidth]{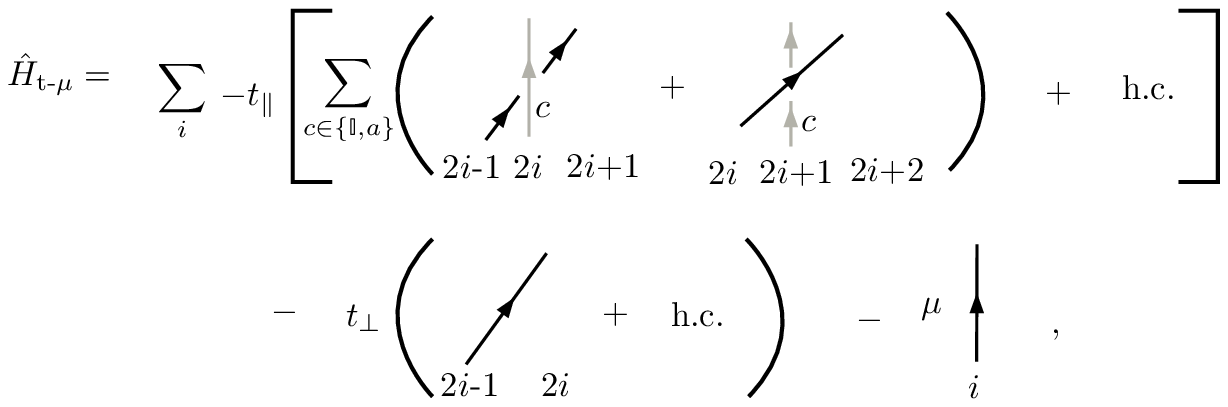}
 \label{KineticTerm}
\end{equation}
where $t_{\parallel}$ is the horizontal hopping strength on the legs, $t_{\perp}$ is the transverse hopping strength on the rungs, and $\mu$ is the chemical potential. The term ``h.c.'' denotes the Hermitian conjugate of all the diagrammatic terms preceding it. In our diagrammatic expression, a solid black line represents a non-trivial charge $a$, which can be $\tau$ or $\sigma$ depending on the model, while a grey line represents either the trivial vacuum charge $\mathbb{I}$ or a nontrivial charge $a$. 

Note that, when $c = \mathbb{I}$, hopping does not introduce any braid factor, but when $c=a$, the hopping term acquires a braid factor $R^{aa}$ for ``under-crossing,'' and $(R^{aa})^{\dagger}$ for ``over-crossing.'' The matrix representation of these braid factors, for both Fibonacci and Ising anyons, are given in Appendix~\ref{App:FusionData}.

\subsubsection{Heisenberg interaction term}
The interaction term is a sum of Heisenberg-like couplings applied to anyons. For two non-trivial anyonic charges with label $a$, and fusion outcomes $a \times a \rightarrow \mathbb{I} + \cdots$, an antiferromagnetic coupling energetically favours projection of the two charges $a$ into the vacuum channel $\mathbb{I}$, being the process with lowest energy cost. Ferromagnetic couplings between particles would have preferred the other fusion outcome.

With reference to Fig.~\ref{FusionOrder}(i), we project the non-trivial charges $a$ on sites $(2i-1, 2i+1)$ on the top leg and sites $(2i, 2i+2)$ on the bottom leg to the vacuum charge $\mathbb{I}$ with an interaction strength $J_{\parallel}$, and also project the charges $a$ on sites $(2i-1, 2i)$ on the rung to the vacuum charge $\mathbb{I}$ with an interaction strength $J_{\perp}$. Diagrammatically, the interaction Hamiltonian is,
\begin{equation}
 \includegraphics[width=\linewidth]{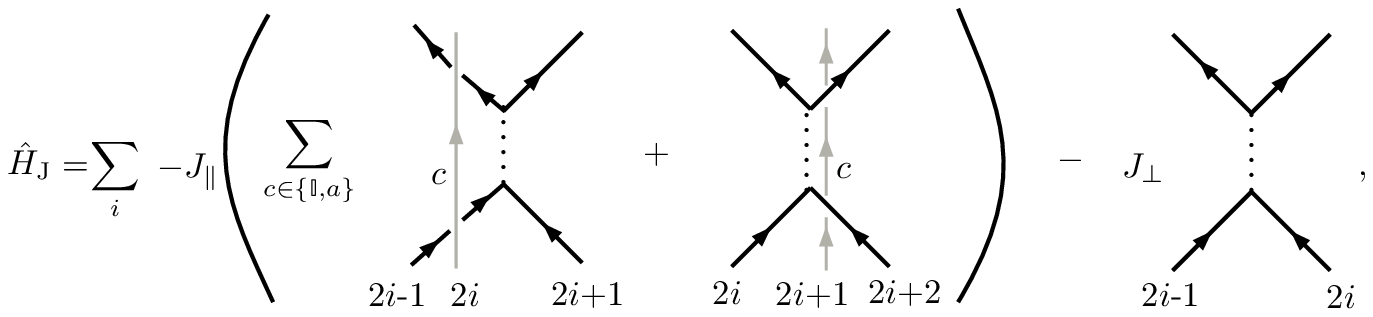}
\label{InteractionTerm}
\end{equation}
where the dotted line represents the vacuum charge $\mathbb{I}$. For this work we focus on the fully antiferromagnetic interaction regime, i.e. $J_{\parallel}, J_{\perp} = 1$.

% Presentation of results and discussion
\section{Results}
\label{Sec:Results}
In this section we present the phase diagrams of our model for both Fibonacci and Ising anyons. Our results are based on numerical analyses using \emph{anyonic} MPS as the Ans\"atze, and \emph{anyonic} TEBD and \emph{anyonic} DMRG as the algorithms. The number of basis states kept in the MPS (i.e. bond dimension) is $\chi = 200$.
 
The Hamiltonian defined in Eq.~\ref{KineticTerm} and \ref{InteractionTerm} have several parameters: hopping term $t_{\perp}$ of particles hopping on the rungs (or simply ``rung hopping''), hopping term $t_{\parallel}$ on legs  (or ``leg hopping''), a chemical potential term $\mu$ that controls the filling density of the ladder, and Heisenberg interaction terms on the rungs $J_{\perp}$ (or ``rung interaction'') and on the legs $J_{\parallel}$ (or ``leg interaction''). This makes the parameter space very large, so we make some simplifications. Our goal is to understand the effect of braid statistics on the ground states of anyonic systems, and therefore the leg hopping $t_{\parallel}$ and the chemical potential $\mu$ are sufficient to ``tune'' the effect of braidings of anyons. Hence, we make only those  two parameters tunable, while the rest are fixed. In order to maintain a ladder model, we set $t_{\perp} = 1$---i.e. we do not consider an instance of having two copies of a single chain (when $t_{\perp}=0$). The ground space of \emph{free} non-Abelian anyons is exponentially degenerate in the number of anyons in the system, which becomes lifted in the presence of inter-particle interactions. Henceforth, we fix the interaction strengths to $J_{\parallel}=J_{\perp}\equiv J=1$ (i.e. both antiferromagnetic on the legs and rungs of the ladder). 

The order parameters we used in detecting phase transitions in our model are the  average particle density per rung,  $\nu = \langle \hat{n} \rangle$, where $\hat{n}$ is the particle number operator, and the block scaling of the entanglement entropy defined as, 
\begin{equation}\label{Eq:EE1}
S(l) = -\mathrm{Tr} \left(\hat{\rho}_l ~\mathrm{log} ~\hat{\rho}_l \right),
\end{equation}
where $\hat{\rho}_l$ is the reduced density matrix of a contiguous block of $l$  sites (i.e. coarse-grained rungs) on the ladder. 

For critical systems with open boundaries and in the thermodynamic limit, the entanglement entropy has a simple relation from conformal field theory (CFT),\citep{Cardy1996, Francesco1997, Calabrese2009a}
\begin{equation}\label{Eq:EE2}
 S(l) = \frac{c}{3} ~ \log ~l,
\end{equation}
where $c$ is the central charge of the CFT. Hence at criticality we can calculate $c$ and map the ground state to an appropriate CFT. The central charge can be computed from the slope, which may be approximated from the numerical data using the formula
\begin{equation}\label{Eq:EquationForCentralCharge}
c = 3\times \left( \frac{S(l_2) - S(l_1)}{\log l_2 -\log l_1} \right),
\end{equation}
where $l_1$ and $l_2$ are two different sizes of the block of sites chosen for the calculation. 

\begin{figure}
\subfloat[Fibonacci anyons]{%
  \includegraphics[clip,width=\columnwidth]{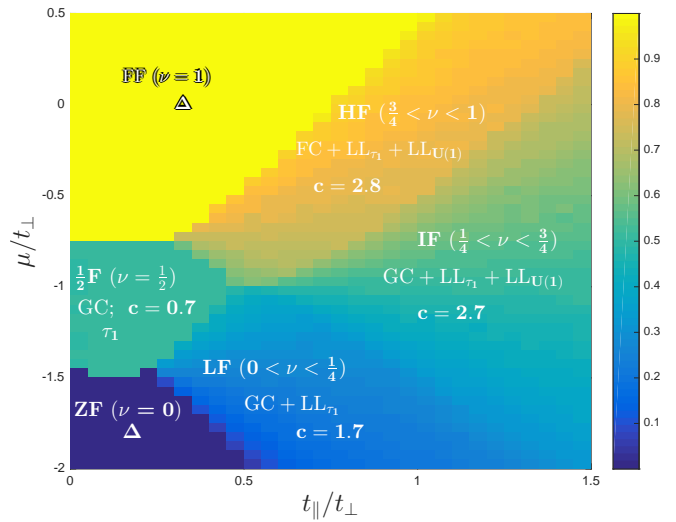}%
} \\
\subfloat[Ising anyons]{%
  \includegraphics[clip,width=\columnwidth]{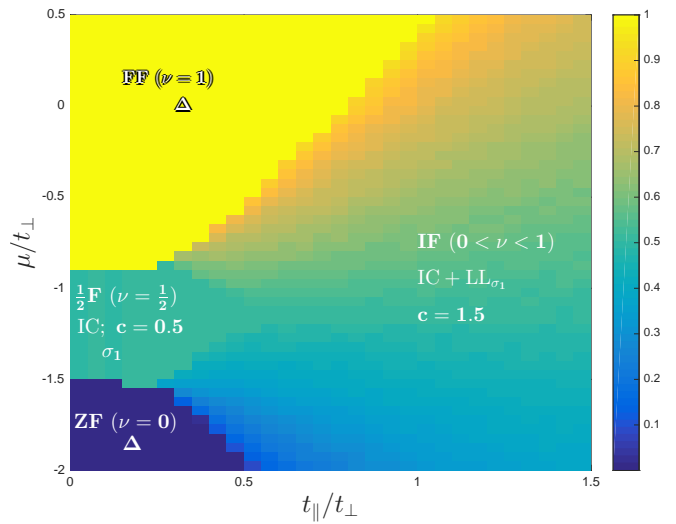}%
}
\caption{Phase diagrams of both anyon models: $(a)$ Fibonacci and $(b)$ Ising anyons. The horizontal axis is the leg hopping ratio $t_{\parallel}/t_{\perp}$, the vertical axis is the chemical potential $\mu/t_{\perp}$, and the plotted quantity is the filling fraction $\nu=\langle \hat{n} \rangle$. Six distinct phases are apparent for Fibonacci anyons, and four for Ising anyons. We describe these phases in terms of the properties of their ground states, including the nature of filling and average particle density in (round brackets), the central charge $c$ for gapless phases, and the types of spin and charge excitations in the critical phases. For convenience of reference, we assign some names to distinct sectors of  the phase diagrams based on the following mnemonics: ``ZF'' for zero filling ($\nu=0$), ``$\frac{1}{2}$F'' for half-filling ($\nu = 1/2$) at low leg hopping, ``FF'' for fully filled ($\nu=1$), ``LF'' for low filling ($0 < \nu < \frac{1}{4}$), ``IF'' for intermediate filling ($\frac{1}{4} < \nu < \frac{3}{4}$), ``HF'' for high filling ($\frac{3}{4} < \nu < 1$), ``$\Delta$''for a gapped phase, ``GC'' for \emph{effective} Golden chain,\cite{Feiguin2007} ``IC'' for \emph{effective} Ising chain, ``FC'' for \emph{effective} ``Fibonacci crystal'' structure,  ``$\mathrm{LL}_{\tau_1}$'' for a Luttinger liquid of hopping $\tau_1$ charges, ``$\mathrm{LL}_{\sigma_1}$'' for a Luttinger liquid of hopping $\sigma_1$ charges, and ``$\mathrm{LL}_{U(1)}$'' for a Luttinger liquid of hopping bosonic U(1) charges. Some of the terms which are unfamiliar are explained in the text.}
\label{Fig:SurfacePhaseDiagrams}
\end{figure}

\begin{figure}
\subfloat[Fibonacci anyons]{%
  \includegraphics[clip,width=\columnwidth]{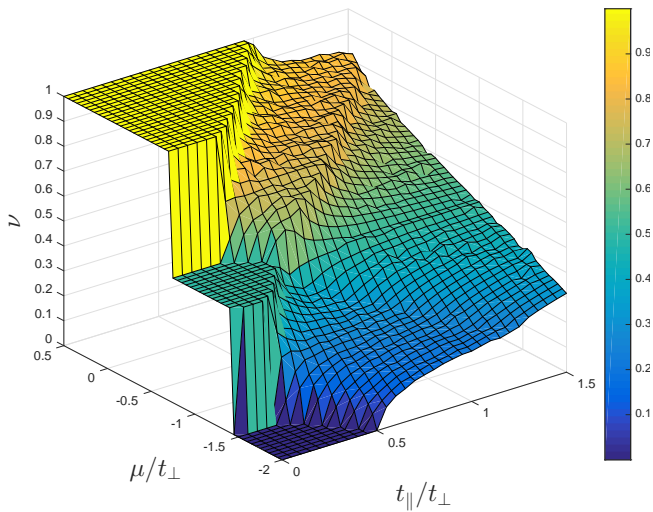}%
} \\
\subfloat[Ising anyons]{%
  \includegraphics[clip,width=\columnwidth]{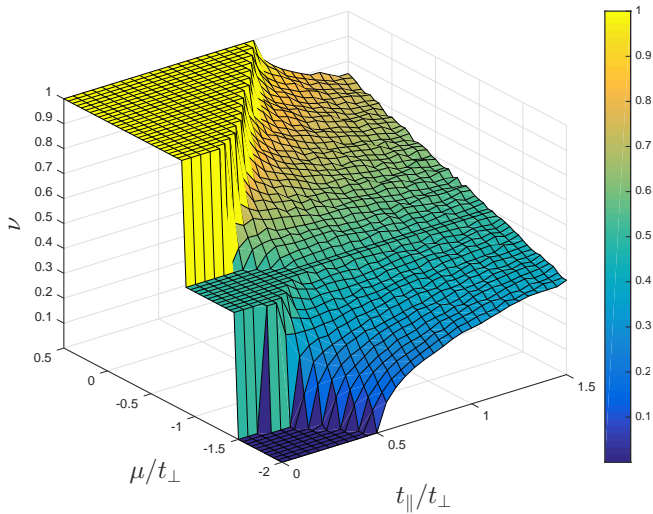} }
\caption{Three-dimensional versions of Fig.~\ref{Fig:SurfacePhaseDiagrams}. The vertical axis is the filling fraction---which was collapsed in Fig.~\ref{Fig:SurfacePhaseDiagrams}}
\label{Fig:3DPhaseDiagrams}
\end{figure}

% Present phase diagrams here
The phase diagrams of our model are presented in Fig.~\ref{Fig:SurfacePhaseDiagrams} (planar view) and Fig.~\ref{Fig:3DPhaseDiagrams} (three dimensional view). In most of the discussion, we will use Fig.~\ref{Fig:SurfacePhaseDiagrams} to explain the physics in the ground states of our model as the control parameters change. The vertical axis is the chemical potential $\mu/t_{\perp}$ and the horizontal axis is the leg hopping amplitude $t_{\parallel}/t_{\perp}$. 

As the microscopic couplings of both anyon models are \emph{similar}, we attribute the differences in the phase diagrams to the differing fusion and braid statistics of Fibonacci and Ising anyons.

We identified two gapped and four critical phases in Fibonacci anyons, and two gapped and two critical phases in Ising anyons. In the Fibonacci anyons phase diagram Fig.~\ref{Fig:SurfacePhaseDiagrams}(a), the four critical distinct phases are labelled as ``$\frac{1}{2}$F,'' ``LF,'' ``IF,'' and ``HF.'' The two distinct critical phases in the Ising anyons are also labelled as ``$\frac{1}{2}$F'' and ``IF.'' The meaning of these mnemonics will be explained later. 

\begin{figure}
\includegraphics[width=\columnwidth]{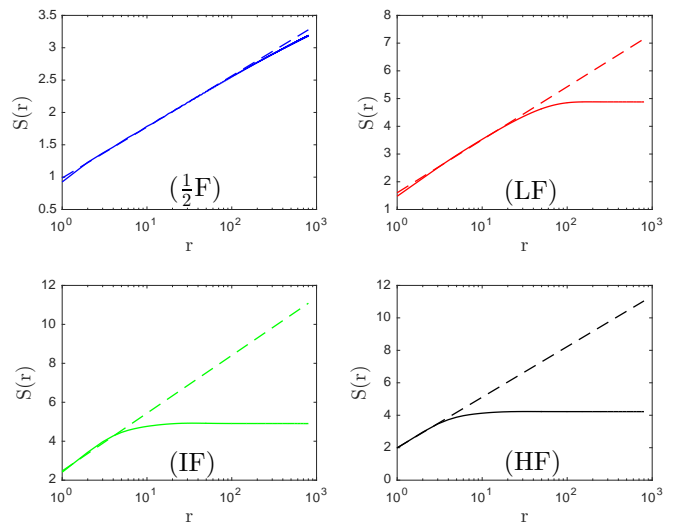}
\caption{Plots of block scaling of the entanglement entropy for Fibonacci anyons. Individual subplot belongs to a distinct phase in the Fibonacci anyons phase diagram Fig.~\ref{Fig:SurfacePhaseDiagrams}(a). The same set of labels as used in Fig.~\ref{Fig:SurfacePhaseDiagrams}(a) are also used here for easy identification. These plots are for some representative states within the phases at  some value of $(\mu, t_{\parallel})$. For $\frac{1}{2}$F, $(-1, 0.25)$; for LF, $(-1.5, 0.75)$; for IF, $(-1, 1.5)$; and for HF, $(-0.45, 0.75)$. The solid lines are the data, and the dashed lines are the linear fit to the  plots. The saturation seen in the plots is the effect of the limited bond dimension of $\chi=200$ used to capture a critical state in the MPS; they are not indications of a gapped phase in this case.}
\label{Fig:EEFibPhases}
\end{figure}

\begin{figure}
\includegraphics[width=\columnwidth]{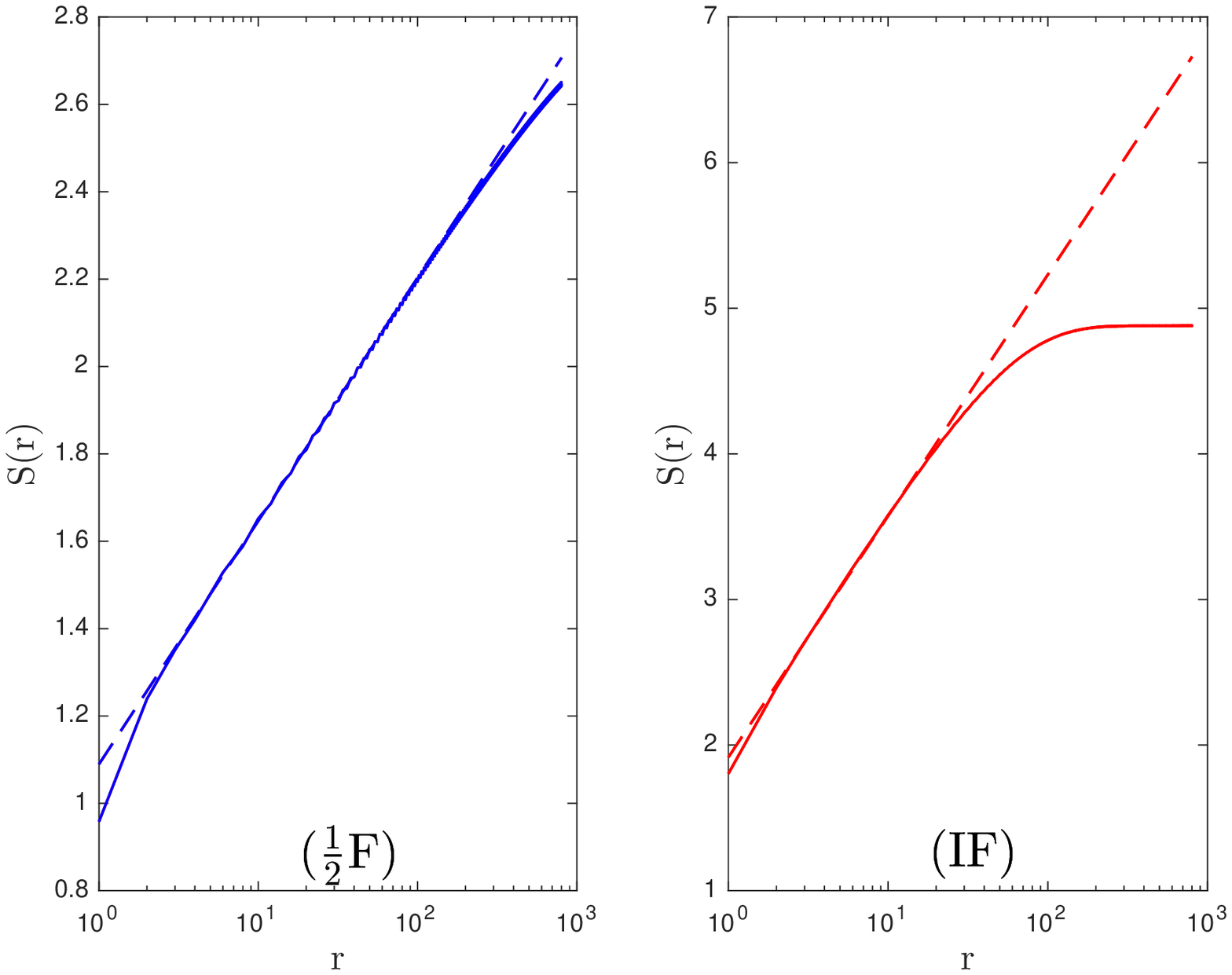}
\caption{Plots of block scaling of the entanglement entropy for  Ising anyons. Individual subplot belongs to a distinct phase in the Ising anyons phase diagram Fig.~\ref{Fig:SurfacePhaseDiagrams}(b). The same set of labels as used in Fig.~\ref{Fig:SurfacePhaseDiagrams}(b) are also used here for easy identification. These plots are for some representative states within the phases at  some value of $(\mu, t_{\parallel})$. For $\frac{1}{2}$F, $(-1, 0.25)$; and for  IF, $(-1, 1)$. The solid lines are the data, and the dashed lines are the linear fit to the  plots. The saturation is the effect of finite bond dimesion of $\chi=200$ used in the MPS.}
\label{Fig:EEIsingPhases}
\end{figure}

The plots of the block scaling of the entanglement entropy of the critical phases in Fibonacci and Ising anyons are given in Fig.~\ref{Fig:EEFibPhases} and Fig.~\ref{Fig:EEIsingPhases}  respectively.

We computed the value of the central charges of the critical phases of both anyon models from the slope of the block scaling of the entanglement entropy $S(l)$ for an increasing block size $l$ using Eq.~\ref{Eq:EE2} and Eq.~\ref{Eq:EquationForCentralCharge}. These central charges are given in Table~\ref{Tab:CentralChargesTable}

\begin{table}
\begin{ruledtabular}
\begin{tabular}{  c |  c   | c | c | c  }
Phases   & $\frac{1}{2}$F & LF & IF & HF  \\[0.5pt] \hline

Fibonacci anyons:\quad $c_{\mathrm{num}}$  &   $0.71263$  & $1.7259$ & $2.6786$ & $2.8146 $ \\[0.5pt] \hline

Ising anyons: \quad $c_{\mathrm{num}}$  &   $0.503$  &  --- & $1.4967$ & --- \\
\end{tabular}
\end{ruledtabular}
\caption{The numerical values of the central charges $c_{\mathrm{num}}$ of the critical phases in Fibonacci anyons (in the second row) and Ising anyons (in the third row). These values are extracted from the slope of the block scaling of the entanglement entropy $S(l)$ for an increasing block size $l$.} 
\label{Tab:CentralChargesTable}
\end{table}

Ideally, the block scaling of the entanglement entropy $S(l)$ of critical states for a block size of $l$ should be a straight line in logarithmic scale according to  Eq.~\ref{Eq:EE2} from CFT. In numerical simulations of critical states using infinite MPS with a finite bond dimension $\chi$, the expected linearity is, in practice, not always reproduced if there are lots of entanglement in the system. Rather, what is observed is that after some block sizes $l_*$, the value of $S(l_*)$ saturates to some constant value. If by increasing the value of $\chi$, the saturation point $l_*$ also increases (i.e. the linearity of $S(l)$ extends further), that may imply that in the limit that $\chi \rightarrow \infty $, the plot of $S(l)$ against $l$ will be a straight line for $l \rightarrow \infty$. This is a numerical indication of a critical state. The central charge of the theory can then be reliably obtained from the slope of $S(l)$ using the discrete approximation Eq.~\ref{Eq:EE2}. This procedure is referred to as \emph{finite-entanglement analysis} or \emph{finite-$\chi$ scaling}.\cite{Tagliacozzo2008, Pollman2009} The central charges presented in Table.~\ref{Tab:CentralChargesTable} are obtained from the slope of $S(l)$ after doing finite-entanglement analysis for bond dimensions of $\chi = \{100, 150, 200, 250, 300 \}$ (not shown). 

We discuss all these results in the next section.

% Discussion
\section{Discussion}
\label{Sec:Discussion}
While the phase diagrams of Fig.~\ref{Fig:SurfacePhaseDiagrams} of both anyon models look different at high leg hopping ratio $t_{\parallel}/t_{\perp}$, they otherwise look similar at low leg hopping. Therefore, we divide the explanation of the phase diagrams into two subsections: one for low leg hopping ratio and the other for high leg hopping ratio.

\subsection{Low hopping ratio} \label{Sec:LowHopping}
% First-hand general overview of the phase diagram
We will regard the value of the leg hopping amplitude to be ``low'' if it is not sufficient to make the system become superfluid. This value also depends on the chemical potential $\mu$. Beyond some critical value of $t_{\parallel}/t_{\perp}$, a phase transition occurs, and the system becomes superfluid. At low leg hoppings, the phase diagrams of the two anyon model look similar, and can hence be given a unified explanation.

The chemical potential enters the Hamiltonian as $-\mu \hat{n}_i$, where $\hat{n}$ is the particle density operator on a single rung with particle numbers $n_i = \{0,1,2\}$. This means that, on a single rung, particles are added in discrete amounts, from no particle to a maximum of two particles. As the chemical potential varies at low leg hoppings, the system experience a first order phase transition between three distinct phases with different average particle densities. First, from a phase with an average particle density of zero to a phase which is half-filled (i.e. with an average of one particle per rung), and finally to another phase where the ladder is 100 per cent filled (i.e. with an average of two particles per rung ). It is convenient to associate some mnemonics to these phases for ease of reference: the phase with zero filling  is given ``ZF,'' the phase at half filling (but at low leg hopping) is given ``$\frac{1}{2}$F,''  and the phase with full filling is given ``FF.'' All these names are superscribed in their respective sections on the phase diagrams. The quantum phase transitions between these phases are sharply revealed by ``jumps'' in the filling fraction as a function of the chemical potential as shown in Fig.~\ref{Fig:3DPhaseDiagrams}, which is a signature of first-order phase transitions. Next, we explain the critical nature of these three phases. 

% Explanation of the criticality of the phases
At very low values of the chemical potential $\mu$, the system has a negligible number of particles and the particle density averages to zero. This phase is trivially gapped ($\Delta$) with the dominant occupation being the trivial vacuum charge $\mathbb{I}_0$ on all sites of the ladder.

By increasing the chemical potential, there is phase transition at a particular critical value. At the critical value (and beyond) particles are added to the ladder, with an average of one particular per rung. A single particle on any rung will be a superposition of being either on the top or bottom site of that rung (since there is a non-zero value of rung hopping. The value of $t_{\perp}$ is fixed to one). At low leg hopping, i.e. $t_{\parallel} < t_{\perp}$, particles will generally prefer to hop on rungs than to hop across, since hopping on rungs have a lower energy cost than hopping across. The particles therefore arrange themselves into a Mott insulator-like state. This opens a charge gap in the charge sector of the ground state. However, the ``anyon sector'' (also called spin sector) in analogy with a similar concept for bosons and fermions, is not gapped, as will be shown. As particles hop mostly on rungs, the nearest neighbouring antiferromagnetic interaction $J_{\parallel}$ on the legs prefer to have two anyons as nearest neighbours, as that lowers the energy. For instance, we would expect that in the spin sector,
\begin{equation}
\begin{matrix}
\includegraphics[scale=1]{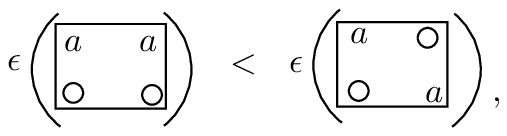}
\end{matrix}
\end{equation} 
where $a$ is either $\tau$ (Fibonacci anyon) or $\sigma$ (Ising anyon). That is, in the spin sector, the local energy $\epsilon(\bullet)$ of the configuration on the left should be less than that on the right on any plaquette. Therefore, in the spin sector particles would prefer to align into chainlike configurations, since that lowers the energy.

The ground state of chains of non-Abelian anyons have been studied.  In particular, it has been shown that a chain of non-Abelian anyons with an Heisenberg-type interaction map to known conformal field theories. In Ref.~\onlinecite{Feiguin2007}, the authors showed that the CFT of a  chain of Fibonacci anyons [also known as the Golden chain (GC)]  has a central charge of $c_{\mathrm{theory}}=0.7$ for pairwise antiferromagnetic couplings, and the value of $c_{\mathrm{theory}}=0.8$ for ferromagnetic couplings. A similar lattice chain, but with Ising anyons has a central charge of $c_{\mathrm{theory}}=0.5$ for both types of couplings. Therefore, while we expect the charge sector to be gapped, the spin sector is gapless and described by CFT.

From our numerical calculations of the block scaling of entanglement entropy at half filling, we found the ground state of both Fibonacci and Ising anyons to be critical. Using Eq.~\ref{Eq:EquationForCentralCharge}, we computed the central charges of their CFT to be, $c_{\mathrm{num}}=0.7029$ for Fibonacci anyons, and $c_{\mathrm{num}} = 0.499$ for Ising anyons, which are both very close to their theoretical values. These show that the spin sector is indeed gapless, and therefore supports our conjecture that at half filling and low hopping amplitudes, particles organize into chainlike configurations.

Finally, as the chemical potential increases further, there is a final transition into a phase where the system becomes 100 per cent filled. We note that, there is no other phase transition into a $(\nu - \epsilon)$-filling phase, where $\epsilon$ is a small number around the observed densities of $\nu = 0, 0.5$ and $1$. We give one qualitative reason for the gapped ground state behaviour at higher values of the chemical potential. Consider the value of $\mu/t_{\perp}=0$ for example (the average particle density to be $\nu=1$). There remains only two competing interactions in the Hamiltonian, namely hopping and interaction strengths on the rungs and legs of the ladder, where all couplings are equal and set to $1$, except the horizontal leg hopping $t_{\parallel}/t_{\perp}<1$. As there can be either $0$, $1$, or $2$ particles per rung, we wish to determine which occupation gives the state with the lowest energy. The energy for a single particle to hop on one rung is $-t_{\perp}$, which is the same as the interaction energy $-J_{\perp}$ for coupling two particles on that rung, and so, there is no preference for having either one or two particles on that rung. However, as $J_{\parallel}>t_{\parallel}$, it is energetically favourable to place two neighbouring particles on the top and bottom legs of the ladder. Therefore, the ground state prefers to have two nearest neighbouring particles on each leg of the ladder, and is hence fully filled.

The fully filled (FF) phase is gapped. It has been shown in previous works~\cite{Poilblanc2011, Ayeni2016} that a two-leg ladder which is fully filled with either Fibonacci or Ising anyons is gapped. This can be simply understood. There is a strong vertical AFM Heisenberg interaction $J_{\perp}=1$ that favours fusing pair of charges on every rung into the vacuum charge, which is a product state under renormalization. Our calculations reconfirmed this.

We next turn to the explanation of the phase diagrams at high leg hopping ratio.

\subsection{High hopping ratio} \label{Sec:HighHopping}
When the leg hopping $t_{\parallel}/t_{\perp}$ increased beyond a critical value, for all values of the chemical potential, we observed second-order phase transitions for both anyon models (see Fig.~\ref{Fig:SurfacePhaseDiagrams} and Fig.~\ref{Fig:3DPhaseDiagrams}). We found the new phases to be superfluid (as will be explained below). We will refer to values of the leg hopping $t_{\parallel}/t_{\perp}$ which makes the system superfluid as ``high'' leg  hopping. Essentially, depending on the value of the chemical potential, there are different critical values of the leg hopping ratio $t_{\parallel}/t_{\perp}$.

At high leg hopping, there are three distinct phases for Fibonacci anyons, while there is just only one for Ising anyons. The immediate conclusion that can be drawn from this is that the fusion and braid statistics of these particles are responsible for the observed differences. 

From the phase diagrams (Fig.~\ref{Fig:SurfacePhaseDiagrams}), it can be seen that as the chemical potential $\mu$ varies continuously, the average particle density $\nu$ changes discontinuously across the three distinct phases observed for Fibonacci anyons, but there is no discernible discontinuity in the average density of Ising anyons, even though the range of computed particle density $\nu$ is the same for both models except for the singular points which are not defined in the Fibonacci anyons. We now explain the physics behind these differences.

% Low chemical potential
\subsubsection{Low $\mu$}
At very low values of the chemical potential, that ladder is sparsely populated with an average particle density of $0 < \nu < 1/4$. Therefore, any plaquette on the ladder can have at most one particle ($\tau$ or $\sigma$). As aforementioned, at low leg hopping $t_{\parallel}$, the strong rung hopping $t_{\perp}=1$ opens a gap in the charge sector (as $t_{\parallel} < t_c < t_{\perp}$, where $t_c$ is some critical value). It should expected that at some value of $t_{\parallel} \ge t_c $,\footnote{The critical value $t_c$ depends on the chemical potential.} the charge gap should close and the charge sector becomes gapless. Like the hopping of free bosons or fermions with a central charge of $c=1.0$, the central charge of hopping $\tau$ or $\sigma$ particles is $c=1.0$. In addition, due to the antiferromagnetic couplings favouring fusion of neighbouring particles to the vacuum charge, the particles in the spin sector prefer to arrange themselves into chainlike configurations. As it was shown earlier, a chain of interacting Fibonacci and Ising anyons is critical, and have central charges of $c=0.7$ (antiferromagnetic couplings) and $c=0.5$, respectively. Based on the known phenomenology of spin-charge separation in a chain of itinerant and interacting non-Abelian anyons,\cite{Poilblanc2013} the theoretical values of the central charge of the CFT that describe our ladder model at $0<\nu<1/4$ should be $c_{\mathrm{theory}}=1.7$ and $c_{\mathrm{theory}}=1.5$ for Fibonacci and Ising anyons respectively.

To confirm this prediction, we chose some representative states satisfying the density constraint of $0<\nu<1/4$ for both Fibonacci and Ising anyons to compute the value of $c$. From our calculations, we obtained $c_{\mathrm{num}} = 1.7259$ for Fibonacci anyons, and $c_{\mathrm{num}} = 1.4967$ for Ising anyons. (We computed central charge values of almost the same quality for other chosen states, and is therefore true for all the set of states satisfying the density constraint at high leg hopping). The numerical values are close to the predicted theoretical values and hence support our theoretical analyses.

In the Fibonacci anyon model, we call the set of states with average particle density $0 < \nu < 1/4$ collectively as ``low filling'' (LF) phase (because they form a distinct phase). In the phase diagram Fig.~\ref{Fig:SurfacePhaseDiagrams}(a) we label the low filling phase as follows: (1). LF ($0<\nu<1/4$) specifying the nature of filling and range of particle densities. (2). $\mathrm{GC} + \mathrm{LL_{\tau_1}}$ to indicate that there is a single effective golden chain and one Luttinger liquid of hopping $\tau$ charges. (3). And lastly, the phase is labelled with the value of its central charge, $c = 1.7$. 

In the Ising anyons phase, the set of states with $0<\nu<1/4$ do not form a distinct phase from other states with $\nu \ge 1/4$ (as will be shown). We therefore label all the set of states with $0 < \nu < 1$ as: 1. IF $(0 < \nu < 1)$ where ``IF'' stands for ``intermediate filling.'' 2. $\mathrm{IC} + \mathrm{LL_{\sigma_1}}$ to indicate that in the ground state there can be either a single interacting Ising anyons chain for $0 < \nu < 3/4$ or some nontrivial crystal structure at $3/4 \le \nu < 1$, and one Luttinger liquid of hopping $\sigma$ charges. 3. And lastly, the phase is also labelled with the value of its central charge, $c = 1.5$.

% Moderate value of the chemical potential 
\subsubsection{Moderate $\mu$}
As the chemical potential increases, there is no indication of any other phase transition in the system of Ising anyons. As a result, we group the set of states for all values of the chemical potential $\mu$ that realize the particle densities $0 < \nu < 1$ into one phase, and referred to them collectively as ``intermediate filling.'' In the system of Fibonacci anyons on the other hand, we observed other phase transitions. The set of possible states of Fibonacci anyons at high leg hopping amplitudes group into three  distinct phases: the LF phase (that was previously mentioned), and two new distinct phases which are classified as ``IF'' for ``intermediate filling'' (i.e. the set of states with $1/4 < \nu < 3/4$) and ``HF'' for ``high filling'' (i.e. the set of states with $3/4 < \nu < 1$). We now explain the physics behind the appearance of the two new phases in Fibonacci anyons but which do not manifest in Ising anyons.

We begin by considering filling density of $1/4 < \nu < 3/4$ for some range of values of the chemical potential $\mu$. At these densities the rungs of the ladder can have either one or two particles. In the charge sector, the kinetic term in the Hamiltonian favours the hopping of the ``fundamental'' $\tau$ or $\sigma$ charges against a background of holes. This account for one mobile excitation in both anyon model. But in addition, because of the fusion and braid statistics of the anyon models and restricted dimensionality of the ladder, there is a new \emph{emergent} bosonic charge degree of freedom possible with Fibonacci anyons but not with Ising anyons. The new charge excitation manifest as the hopping of U(1) charges on a background of $\tau$ charges. We show how this happens. For example, consider the effect of the hopping of a single $\tau$ charge on a plaquette with $3$ Fibonacci anyons (represented as occupied sites; vacuum charges as vacant sites): 
\begin{equation*}
\begin{matrix}
\includegraphics[scale=1]{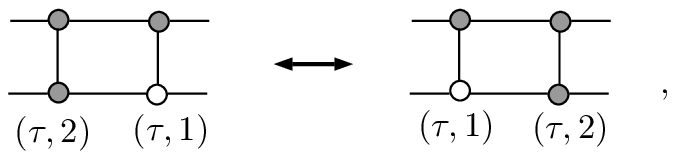}
\end{matrix}
\end{equation*}
where $(\tau,2)$ means that there are two $\tau$ charges which can possibly fuse into another $\tau$ charge, and $(\tau,1)$ which is equivalent to a single $\tau$. When the  $\tau$ charge on the bottom leg hops across, there is an exchange of U(1) charge labels, but the $\tau$ labels do not change. In a ``loose'' term, the exchange process is ``blind'' to the hopping of the $\tau$ charge but aware of the exchange of the ``$1$'' and ``$2$'' U(1) charge labels on the $\tau$ charges. This is effectively another kind of mobile degree of freedom which is different from the hopping of $\tau$ against a background of vacuum charge $\mathbb{I}$. 

It is worth noting that this is not possible with Ising anyons because of its fusion statistics: $\sigma \times \sigma \rightarrow \mathbb{I} + \psi$. A pair of $\sigma$ charges on a rung does not fuse into $(\sigma,2)$, and therefore, there cannot be a ``blind'' exchange of U(1) charge labels on a background of $\sigma$ charges (on the ladder with nearest neighbour hoppings).\footnote{Such a possibility may exist with Ising anyons on a ladder with an odd number $k$ of legs, where $k>1$, as an odd number of $\sigma$ charges can fuse into the $\sigma$ charge sector.}

The two mobile charge excitations identified in Fibonacci anyons  should give a theory in the charge sector with a central charge of $2$, while we still expect $c=1$ for Ising anyons. 

In the spin sector, on the other hand, because of the incomplete filling, the particles prefer to arrange into chainlike configurations. In particular, as $\nu < 3/4$, the configuration which minimizes the ground state energy is effectively chainlike. Later, we show that this no longer holds when $\nu \ge 3/4$, where we conjecture that at $1 > \nu \ge 3/4$, particles arrange into some nontrivial crystal lattice configurations, which looks perhaps like a comb.

We therefore predict that the central charge of the theory at ``intermediate filling'' ($1/4 < \nu < 3/4$) should be $c_{\mathrm{theory}}=2.7$ for Fibonacci anyons, but still $c_{\mathrm{theory}} = 1.5$ for Ising anyons. Our numerics indeed confirmed these predictions. For some representative states within the IF phase in Fibonacci anyons and in the $1/4 <  \nu < 3/4$ regime in the Ising anyon phase diagram, we obtained central charges of $c_{\mathrm{num}} = 2.6786$ (Fibonacci) and $c_{\mathrm{num}} = 1.4967$ (Ising). [These numbers are obtained from very converged simulations at different bond dimensions $\chi=\{100, 150, 200, 250, 300 \}$ which are used for finite-entanglement analysis (not shown)]. 

Again, the numerical values are very near the predicted theoretical values, thereby validating our theoretical analysis. This indicates that the phase transition observed from the LF phase to the IF phase in the Fibonacci anyons phase diagram is as a result of the new possible bosonic U(1) charge excitation which do not manifest in Ising anyons on a two-leg ladder.

\subsubsection{High $\mu$}
Lastly, we examine the physics of the ground state at high values of the chemical potential, i.e. those values sufficient to increase the average particle density beyond the range of $1/4 < \nu < 3/4$ of the intermediate filing phase. As the chemical potential increases, we found evidence for another phase transition in Fibonacci anyons from the ``intermediate filling'' phase to the ``high filling'' phase [as shown in Fig.~\ref{Fig:SurfacePhaseDiagrams}(a)] where the particle density is $3/4 < \nu < 1$.  As we have shown that there are no other phase transitions in Ising anyons, we restrict the following discussion mainly to Fibonacci anyons. 

To understand this phase transition, we need to examine what happens to the spin and charge sectors of the ground state as the chemical potential increases. In the IF phase, we identified that there is one spin and two charge excitations in the ground state. 

In fact, at any incomplete filling of the ladder---but sufficiently high to allow for pairing of charges on rungs, there are two mobile bosonic charges: (1) The ``fundamental'' $\tau$ charges hopping against holes. (2) The effective hopping of U(1) charges against a background of $\tau$ charges. We provide an explanation. Since the ladder is not completely filled, there are some nearest neighbouring rungs, where one rung has two $\tau$ charges which may be fusing into $(\tau,2)$, and the next rung has only one $\tau$ which can also be written as $(\tau,1)$. The hopping of a single $\tau$ across rungs therefore exchange their U(1) charge labels. This accounts for the U(1) charge excitation. Therefore for all filling densities (i.e. both IF and HF phases) where particles could pair on a rung, there are two charge excitations giving a central charge of $c=2.0$. 

Formerly, we showed that in the spin sector, particles arranged themselves into an \emph{effective} golden chain and  has a central charge of $c = 0.7$ (antiferromagnetic). One may conclude that the spin sector of the HF phase too would prefer the chainlike alignment of particles and give a central charge contribution of $c=0.7$. That would be a false conclusion, otherwise there would not be another phase transition between IF and HF phase, since it would imply that they are described by the same CFT with the same central charge of $c_{\mathrm{theory}}=2.7$ (accounting for all spin and charge excitations). From our numerical calculations (and also with the aid of the  finite-entanglement analysis), we computed the central charge of the HF phase to be $c_{\mathrm{num}} = 2.8146$ for a particular representative point within the phase. One might want to quickly associate the significant offset from the theoretical value of $c=2.7$ to the limitations imposed by the use of MPS with a finite bond dimension, but we show otherwise. The correct theoretical value of the central charge of the HF phase should be $c_{\mathrm{theory}} = 2.8$, with $2.0$ as the contribution of the two charge excitations in the charge sector and $0.8$ as the contribution from the spin sector. This can be expected. On a triangular lattice of Fibonacci anyons, the authors in Ref.~\onlinecite{Trebst2008} also found a phase whose CFT has a similar central charge for (anti)-ferromagnetic couplings of two and three anyon fusion. One way to test our hypothesis would be to use Anyon$\times$U(1) MPS,\cite{Ayeni2016} at say a fixed filling fraction of $\nu = 4/5$, and set the hopping amplitudes to zero, i.e. $\{ t_{\parallel}, t_{\perp} \} = 0$. In other words, ``lock'' the system entirely into the spin sector---since we already know that the spin and charge degrees of freedom separate. From the block scaling of the entanglement entropy $S(l)$, we extracted a very precise value of the central charge as $c \simeq 0.798$. (There was no visible saturation in the plot of $S(l)$ against $l$ associated with finite-$\chi$ for up to the $800$ rung-site which we considered for the calculation). 

There a number of deductions we make from this observation. First, we conjecture that at filling densities $3/4 \le \nu < 1$, particles on the ladder do not actually form effective chainlike configurations, but a nontrivial crystal structure that cannot be converted into a single chain of connected $\tau$ particles with a finite amount of energy. Second, the effective couplings between the particles now behave ferromagnetic even though the ``fundamental'' couplings between the particles is antiferromagnetic. In Appendix~\ref{App:FibLadderAt3by4Filling}, we provide an example of how to reach these deductions using a simple analysis of a ladder of Fibonacci anyons at $\nu=3/4$ filling.

\section*{}

\section{Conclusions}
\label{Sec:Conclusion}
In this work, we have investigated the ground state properties of an anyonic Hubbard model on a ladder for two types of non-Abelian anyons: Fibonacci and Ising. We found distinct phase diagrams for the two anyon models, where their differences arise from their differing fusion and braid statistics. We have shown that at low leg hopping amplitude and at half filling, particles form effective chains, while at high leg hopping amplitude particles separate into distinct charge and spin components in both Ising and Fibonacci anyons, but with the appearance of a new bosonic U(1) charge excitation in Fibonacci anyons which does not manifest in Ising anyons because of its fusion statistics. In addition, the effective nature of the interaction couplings could change because of the dimensionality and filling density of the system.

There are many important questions yet to be answered, such as how the physics changes if all the interaction couplings are ferromagnetic instead of the antiferromagnetic considered in this paper, or perhaps an interpolation between both. Also important is how the physics change with the number of ladder legs, ultimately reaching the infinite two-dimensional limit. Would spin-charge separation still survives for such larger systems and what types of mobile excitations would be possible in the ground state? Because braiding leads to entanglement in the fusion degrees of freedom, which can affect transport properties, \cite{Lehman2010} it would also be of interest to study out-of-equilibrium behaviour of hopping anyons on a lattice.

\section{Acknowledgements}
We thank Sukhwinder Singh and Nathan McMahon for discussions at different stages of this project. B.M.A thanks Ian McCulloch for useful suggestions at the latter end, and for hospitality at the University of Queensland where this project was ultimately completed. This research was supported in part by the ARC Centre of Excellence in Engineered Quantum Systems (EQuS), Project No. CE110001013.

% References
\bibliography{References}

\appendix 

% Fibonacci anyon ladder at three-quarter filling.
\section{Fibonacci anyons ladder at $\nu=3/4$}
\label{App:FibLadderAt3by4Filling}
In this Appendix, we examine the kind of lattice configurations particles form in the ground state, and what their critical properties are. 

Consider a ladder of Fibonacci anyons at a fixed filling density of $\nu=3/4$, i.e. there are $3$ particles and 1 vacant site per plaquette. There is no hopping of particles on the ladder, save only the Heisenberg interaction terms which couple particles on the legs and rungs of the ladder. As in the main text, nearest neighbouring particles interact with an antiferromagnetic coupling, where pairs of $\tau$ particles along any edge fuse into the vacuum charge $\mathbb{I}$. 

The most elementary ``building block'' of a ladder whose filling density is constrained to be $\nu = 3/4$ is a plaquette; this is a square structure with four sites. The uniform filling fraction of $\nu=3/4$ demand that \emph{any} plaquette on the ladder must have $3$ particles and $1$ empty site. We start by enumerating all possible configurations on a single plaquette:
\begin{equation*}
\includegraphics{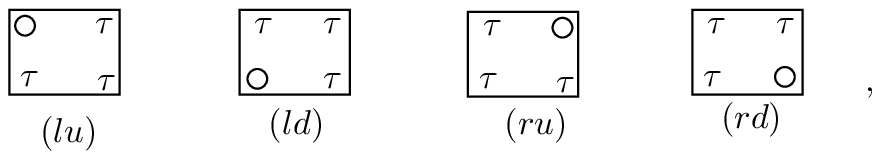}
\end{equation*}
where, for convenience, let the mnemonic in brackets under each plaquette be their name respectively, i.e. $(lu)$ for ``left-up'' (based on the position of the vacant site), $(ld)$ for ``left-down'', $(ru)$ for ``right-up'', and finally $(rd)$ for ``right-down.'' Each of these plaquettes is exactly $3/4$ filled. By composing together \emph{some} of these elementary plaquettes, we can ``build'' a ladder that will have a \emph{uniform} density of $\nu=3/4$.  Here, ``uniform'' is used to mean that any plaquette on the ladder will have an exact filling of $3/4$. This also implies that the ``emerging'' plaquettes that form around the boundaries of the elementary ones should also have a filling density of $3/4$. Which plaquettes from the four enumerated ones might we compose together to give a ladder with a uniform density of $3/4$? The first immediate and easiest choice is to compose individual plaquette with itself. It is also possible to compose some different plaquettes together and still realize the desired system. For example, compose  two copies of $(lu)$ and $(ld)$ together as below:
\begin{equation}
\begin{matrix}
\includegraphics[scale=0.9]{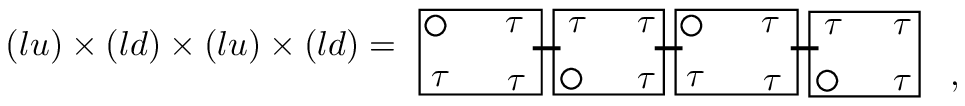}
\end{matrix}
\end{equation}
where the ``$\times$'' represent composition. In this case, we realize a ladder with uniform filling of $3/4$ on any plaquette; either on the elementary plaquettes or the new plaquettes that form around the boundaries of the two elementary ones. Whereas, composing $(lu)$ and $(ru)$ together does not give the uniform $3/4$ filling as shown:
\begin{equation}
\begin{matrix}
\includegraphics[scale=0.9]{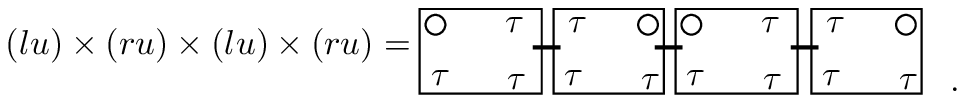}
\end{matrix}
\end{equation}
In this case, there are plaquettes with density which differs from $3/4$, e.g. the plaquette on the boundary of the first $(lu) \times (ru)$ has 100\% filling, while the boundary around the second and third plaquettes has a 50\% filling. The ladder from this type of composition do not give a configuration with a uniform filling of $3/4$, though their average filling is still $3/4$. As such there are specific sets of elementary plaquettes we can compose together to give a uniform $3/4$-filled ladder. Group those sets of plaquettes into: $L = \{(lu), (ld)\}$ and $R = \{(ru), (rd)\}$, where in set $L$, the vacant site is located on the left side of the plaquettes, while in set $R$, the vacant site is located on the right side. Composing elements from within any of the two sets can give the desired system. In other words, either composing elements entirely from set $L$ or from $R$ can give a ladder with a uniform filling of $3/4$ but not mixing elements from both sets. In the rest of this analysis, we can restrict to examining the properties of the ladder built from any of the two sets, say $L$ for the sake of argument, but we essentially reach the same conclusion for set $R$.

Since the particles interact with an antiferromagnetic coupling, which composition minimizes the energy the most? The Fibonacci anyon has the fusion rule $\tau \times \tau \rightarrow \mathbb{I} + \tau$. Let the energy cost of having two $\tau$'s fuse into the vacuum charge $\mathbb{I}$ be $-1$ and let the energy of fusing two $\tau$'s into $\tau$ be $0$. That is, the ``fundamental'' interaction between nearest neighbouring $\tau$'s along any edge (vertical or horizontal) of  the ladder is antiferromagnetic. 

Using set $L$ for illustration, each plaquette from this set have the same energy since each $\tau$ charge interact with at most two other $\tau$'s. But putting the plaquettes together can either lower or increase the energy depending on how we combine them. Composing two similar plaquettes together has a lower energy than composing two dissimilar ones. Therefore, for example, we expect
\begin{equation}
\includegraphics{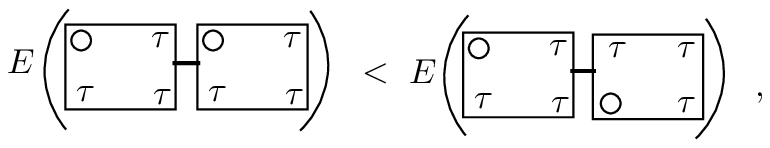}
\end{equation}
also given symbolically as
\begin{equation}
E(lu \times lu) < E(lu \times ld),
\end{equation}
where the $\times$ represent composition. In the composition on the left, particles can interact with at most three other particles, whereas in the composition on the right, particles can interact with at most two other particles, hence the type of composition on the left (i.e. composing same plaquettes) has a lower energy than the type of composition on the right. Based on these heuristics, we can make a general statement here: composing the same plaquettes lower the energy while composing dissimilar ones raise the energy. Therefore, lowest energy configurations would prefer to have two similar plaquettes as nearest neighbours.

Given $N$ elementary plaquettes, there are $2^N$ possible ways to build a ladder of $\nu=3/4$ from either set $L$ or $R$, since there are only two elements in each set. Using set L say, the ladder built using either $(lu)$ or $(ld)$ as $(lu)^{\times N}$ or $(ld)^{\times N}$ have the lowest energy of the possible $2^N$ configurations, and will therefore be parts of the configurations that dominate the ground state. Whereas, configurations having a mix of both $(lu)$ and $(ld)$ composed together have higher energies. 

Having realized compositions which can give the lowest and highest energy, let us invoke yet another classification that would explain their ground state properties better. Let the ladder made out of same plaquettes be classified as ``FC'' and the one made from alternating types of plaquette be  ``GC''. The definitions of these acronym are in order. Schematically, 
\begin{equation}\label{Eq:FC&GC}
%\begin{matrix}
\includegraphics{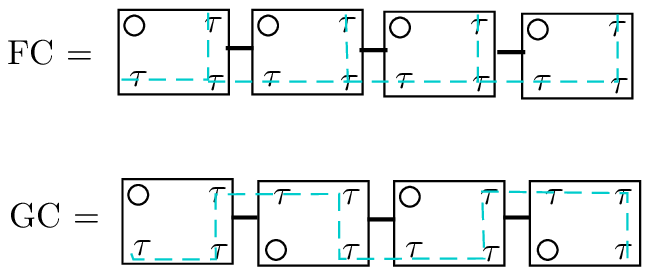}
%\end{matrix}
\end{equation}
It can be observed that while a single continuous line can be traced through GC, no single line can be traced through FC without lifting the hand, not even by any trick. Therefore, GC is effectively a single chain, which has been studied in the literature under the name ``Golden chain'' (hence ``GC''). On the other hand, FC is not equivalent to GC. On an infinite lattice, it would take an infinite amount of deformations to convert FC into GC. At best, the line which minimizes the interaction distance the most in FC would look like a ``comb,'' hence ``FC'' stands for Fibonacci (anyons) comb, alluding to Dirac comb (which is a lattice of potentials of the Dirac delta-function type) in textbook introductory quantum mechanics. The FC and GC are only specific configurations out of the $2^N$ possible configurations.  There are many other configurations but which can be essentially classified as either effective golden chain or ``Fibonacci comb,'' or some nontrivial crystal structure. 

Therefore, in a system of many particles where they are free to choose their own preferred configurations, they will prefer the 		``comblike'' structure more than the chainlike structure.

The ground state properties of GC has been studied, with known ground state energy and central charge. But the FC has not been specifically mentioned. For the specific configurations shown in Eq.~\ref{Eq:FC&GC}, FC should have a lower energy than GC, which we confirmed from our numerics: the energy per particle in the Fibonacci comblike lattice is $\epsilon_{\mathrm{FC}} = -1.12693$, which is lower than the energy per particle in the golden chain, $\epsilon_{\mathrm{GC}} = -0.76393$.\cite{Singh} Results are obtained from anyonic MPS with anyonic TEBD algorithm using a bond dimension of $\chi=200$.

\begin{figure}[ht!]
\includegraphics[width = 0.85\columnwidth]{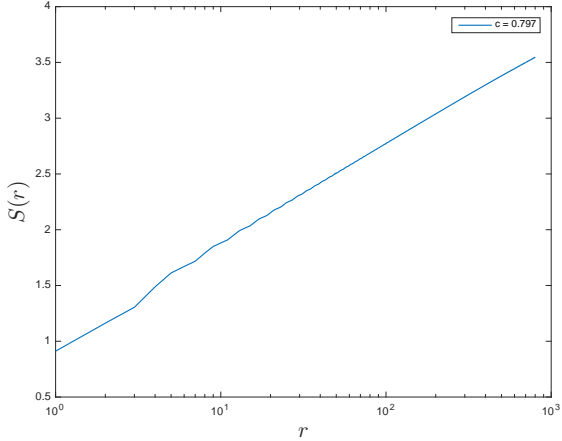}
\label{Fig:FibCombBlockEntropy}
\caption{The block scaling of the entanglement entropy $S(r)$ for the Fibonacci comblike structure in Eq.~\ref{Eq:FC&GC}. No effect of finite bond dimension is visible in this plot up to $800$ sites, the maximum block we considered in computing $S(r)$.}
\end{figure}

Therefore, at filling density $\nu=3/4$ (and higher but less than $1$ ), particles on the ladder would prefer to organize as FC rather than GC in the spin sector. The ground state of GC has a central charge of $c=7/10$ (as the fundamental couplings are antiferromagnetic), but the central charge obtained for FC is $c = 0.797$. The graph of the block entanglement scaling for FC is Fig.~\ref{Fig:FibCombBlockEntropy}

Therefore, though the fundamental interactions are antiferromagnetic, at density $\nu =3/4$ on the ladder, the couplings become effectively ferromagnetic. This further supports the fact that the contribution of the  central charge in the spin sector of the HF in the main text is $c=0.8$ rather than $c=0.7$.

\section*{}

% Anyon algebraic data
\section{Ising and Fibonacci anyon data}
\label{App:FusionData}
\subsection{Ising anyons}
The charges in the Ising anyon theory are: vacuum $\mathbb{I}$, Ising anyon $\sigma$, and fermion $\psi$. The fusion rules satisfy
\begin{align}
&\mathbb{I} \times a = a \quad \text{for } a \in \{\mathbb{I}, \sigma, \psi \},\\
&\sigma \times \sigma \rightarrow \mathbb{I} + \psi, \\
&\psi \times \psi \rightarrow \mathbb{I}. 
\end{align}

The nontrivial F-matrix elements are
\begin{align}
 & \left[F_{\sigma}^{\sigma \sigma \sigma} \right]_{e,f} = 
\begin{bmatrix}
 \frac{1}{\sqrt{2}} && \frac{1}{\sqrt{2}} \\
\frac{1}{\sqrt{2}} && -\frac{1}{\sqrt{2}}
\end{bmatrix}_{e, f}, \\
& \left[F_{\psi}^{\sigma \psi \sigma} \right]_{\sigma \sigma} = \left[F_{\sigma}^{\psi \sigma \psi} \right]_{\sigma \sigma} = -1,
\end{align}
where $e,f = \{ \mathbb{I}, \psi \}$.
For other compatible charges, the F-matrix is trivial,
\begin{equation}\label{Eq:Trivial-FMatrix}
\left( F^{abc}_d \right)_e^f = N_{ab}^e N_{bc}^f N_{ec}^d N_{af}^d,
\end{equation}
where $a, b, c$ are the charges fusing to charge $d$ and, $e$ and $f$ are the intermediate charges of the left and right sides of fusion trees in the F-matrix equation.

The R-matrix $R^{ab}=\oplus_c R^{ab}_c$ is a diagonal matrix describing the unimodular phase acquired when particles $a$ and $b$, which fuse to outcome $c$, braid in counterclockwise sense. The R-matrix elements for Ising model anyons are,
\begin{align}
& R_{\mathbb{I}}^{\sigma \sigma} = e^{-i\frac{\pi}{8}}, \quad R_{\mathbb{\psi}}^{\sigma \sigma} = e^{i\frac{3\pi}{8}} \\
& R_{\sigma}^{\sigma \psi} = R_{\sigma}^{\psi \sigma} = e^{-i\frac{\pi}{2}}, \quad R_{\mathbb{I}}^{\psi \psi} = -1.
\end{align}

The quantum dimensions $d$ of the charges are
\begin{equation}
 d_{\mathbb{I}} = d_{\psi} = 1;   \quad d_{\sigma} = \sqrt{2}. 
\end{equation}

\subsection{Fibonacci anyons}
The Fibonacci anyon model contains two charges: vacuum $\mathbb{I}$ and the Fibonacci charge $\tau$, with fusion rules
\begin{align}
&\mathbb{I} \times a = a \quad \text{for } a \in \{\mathbb{I}, \tau \},\\
&\tau \times \tau \rightarrow \mathbb{I} + \tau.
\end{align}

The nontrivial F-matrix elements are
\begin{equation}
\left[F_{\tau}^{\tau \tau \tau} \right]_{e, f} = 
\begin{bmatrix}
 \phi^{-1} && \phi^{-1/2}  \\
 \phi^{-1/2} && -\phi^{-1}
\end{bmatrix}_{e, f}\\
\end{equation}
where $e,f \in \{\mathbb{I}, \tau \}$, and $\phi = \frac{1 + \sqrt{5}}{2}$ is the Golden Ratio. For other compatible charges with trivial $F$-matrix, the value is given in Eq.~\ref{Eq:Trivial-FMatrix}, with $a, b, c, d, e, f \in \{\mathbb{I}, \tau\}$.

The R-matrix elements for Fibonacci anyons are,
\begin{equation}
R_{\mathbb{I}}^{\tau \tau} = e^{-i\frac{4\pi}{5}}, \quad R_{\mathbb{\tau}}^{\tau \tau} = e^{i\frac{3\pi}{5}}.
\end{equation}

The quantum dimensions $d$ of the charges are
\begin{equation}
 d_{\mathbb{I}} = 1;   \quad d_{\tau} = \phi. 
\end{equation}

\end{document}